\documentclass[sigconf]{acmart}

\usepackage{epigraph}

\usepackage{csquotes}
\usepackage{todonotes}
\usepackage{breakcites}

\definecolor{colorA}{RGB}{114,141,196}  
\definecolor{colorB}{RGB}{214,104,99}   
\definecolor{colorC}{RGB}{109,179,147}  
\definecolor{colorD}{RGB}{199,123,227}  
\definecolor{colorE}{RGB}{255,193,116}  
\definecolor{colorF}{RGB}{179,140,121}  

\setcopyright{acmcopyright}
\copyrightyear{2023}
\acmYear{2023}
\acmDOI{XXXXXXX.XXXXXXX}

\acmConference[Journal]{Submission under review}{Under Review}{2023}
\acmPrice{15.00}
\acmISBN{978-1-4503-XXXX-X/18/06}

\begin{document}

\title{Source Code Comprehension: A Contemporary Definition and Conceptual Model for Empirical Investigation}

\author{Marvin Wyrich}
\orcid{0000-0001-8506-3294}
\affiliation{%
  \institution{Saarland University}
  \city{Saarbrücken}
  \country{Germany}
}
\email{wyrich@cs.uni-saarland.de}

\begin{abstract}

Be it in debugging, testing, code review or, more recently, pair programming with AI assistance: in all these activities, software engineers need to understand source code.
Accordingly, plenty of research is taking place in the field to find out, for example, what makes code easy to understand and which tools can best support developers in their comprehension process.
And while any code comprehension researcher certainly has a rough idea of what they mean when they mention a developer having a good understanding of a piece of code, to date, the research community has not managed to define source code comprehension as a concept.
Instead, in primary research on code comprehension, an implicit definition by task prevails, i.e., code comprehension is what the experimental tasks measure.
This approach has two negative consequences.
First, it makes it difficult to conduct secondary research.
Currently, each code comprehension primary study uses different comprehension tasks and measures, and thus it is not clear whether different studies intend to measure the same construct.
Second, authors of a primary study run into the difficulty of justifying their design decisions without a definition of what they attempt to measure.
An operationalization of an insufficiently described construct occurs, which poses a threat to construct validity.

The task of defining code comprehension considering the theory of the past fifty years is not an easy one. 
Nor is it a task that every author of a primary study must accomplish on their own.
Therefore, this paper constitutes a reference work that defines source code comprehension and presents a conceptual framework in which researchers can anchor their empirical code comprehension research.

\end{abstract}



\maketitle

\section{Introduction}

Let us mentally transport ourselves to New York City for a moment. There, we find the Museum of Modern Art, one of the most renowned museums in the world. What could happen if a piece of printed source code were hanging on a large canvas amidst contemporary artworks? Presumably, the visitors would initially be puzzled by what they see. 

Favian is one of these visitors. He stops and gazes at the code for about half a minute, a duration comparable to the average contemplation of an artwork~\cite{Smith:2017:TimeArt}. At first glance, Favian perceives an abstract, unfamiliar language. However, as he approaches closer, he discovers patterns within the lines of text, a rhythm and structure that resonates with him on a subconscious level. And as he begins to discern a deeper meaning behind the code, Favian joyfully concludes, "Now I understand!" and moves on to the next artwork.

Readers with a software engineering background will be skeptical at this point whether our fictional character Favian really \emph{understood} the code. But why? The notion of understanding can be used very differently by different people and in different contexts. This is because we cannot directly observe how well somebody understood code. Instead, code comprehension is a \emph{construct}, a concept created by humans to classify and assign meaning to observed behaviors.
If such a meaning can be agreed upon within a research community, a construct offers added value.\footnote{ See~\cite{Ralph:2018:Construct,Sjoberg:2022:Construct} for an introduction to constructs with examples from software engineering, such as productivity, programming experience, and program maintainability.}

To date, the research community has not agreed on a meaning of code comprehension. Not because there is evident disagreement, but because we have not yet even entered into discourse about it. Accordingly, there has been a lack of publicly articulated perspectives on what code comprehension can be. And without such a perspective, researchers have neither the means to agree with it nor anything to explicitly dissent from. Essentially, the following two issues arise when researchers of primary studies investigate a construct they cannot define.


The first issue relates to what makes sense when it comes to assessing a developer's understanding of code.
A construct like code comprehension can be operationalized.
Favian's self-assessment of his own understanding could be such an operationalization.
The justification of a particular operationalization depends, among other things, on how one defines code comprehension at the conceptual level.
Based on such a definition, it can then be judged whether \enquote{an operationalization seems intuitively reasonable} (\emph{face validity}) and whether \enquote{an operationalization encompasses all aspects of a construct} (\emph{content validity})~\cite{Ralph:2018:Construct}.
The assessment of face and content validity are the first of several steps in the evaluation of construct validity~\cite{Ralph:2018:Construct}.
A general threat to construct validity is that a definition of the construct is missing or insufficient~\cite{Sjoberg:2022:Construct}: if it is not clear what is being measured, it cannot be assessed whether \emph{it} is being measured validly in terms of face and content validity.
\citet{Sjoberg:2022:Construct} note that in software engineering research, most of the concepts are often not theoretically defined.
\citet{Wyrich:2022:40Years} confirmed this observation within the literature on code comprehension experiments.
Hardly any study defines code comprehension.
As a consequence, in most cases, an operationalization of a previously inadequately described construct occurs.
This then leads to limitations in evaluating construct validity.


A second issue of not having a definition arises in the conduct of secondary research.
There are more diverse options available today than ever before to measure a person's code comprehension performance~\cite{Wyrich:2022:40Years}.
At the same time, there is a lack of research on the comparability of different comprehension measures~\cite{Munoz:2023:Evidence}.
The research community was aware of this lack of comparability as early as 40 years ago, when the first studies examined the comparability of specific task designs~\cite{Cook:1984:PreliminaryCloze,Hall:1986:Cloze} and others examined the comparability of various comprehension measures~\cite{Shneiderman:1977:Measuring,Rajlich:1997:TowardsStandards}.
Not much in the direction of comparability has happened since then, and the few existing studies on the comparability of different comprehension measures overwhelmingly conclude that different common ways of measuring code comprehension do not correlate with each other~\cite{Iselin:1988:Conditional,Binkley:2009:Identifier,Ajami:2018:Syntax,Yeh:2021:IdentifyingConfusion,Fakhoury:2020:Lexical}.
All of this not only potentially leads to uncertainty in the design of new primary studies, but also currently leads to each empirical code comprehension study developing its own methodology for measuring the code comprehension performance of its participants~\cite{Wyrich:2022:40Years}.
What was a neglected issue of comparability of comprehension measures is now becoming a much larger issue of comparability of study results in the face of increasing and barely surveyable publication volumes, as authors of potential secondary research cannot know which primary studies intend to measure the same construct.
To put it in a nutshell: source code comprehension makes no sense except in the light of a definition. 

Therefore, we propose a conceptual definition for source code comprehension in this paper.
Our work comprises both a dictionary-like definition of the term itself, and a conceptual model \emph{explaining} what code comprehension \emph{could be} from an ontological point of view. It is an elaborated perspective that other researchers can agree with or (partially) disagree with. In either case, they benefit from having a reference work in which to anchor their research.

The remainder of the paper is structured as follows. In Section~\ref{sec:history}, we travel through the history of code comprehension theory.
In doing so, we learn about milestones in research that have brought us to our current understanding of what researchers mean by code comprehension.
On this foundation, in Section~\ref{sec:model} we then propose a contemporary definition of the term and provide a conceptual model for different types of empirical code comprehension research.
Section~\ref{sec:examples} illustrates with two case examples how researchers can use the conceptual model to anchor their own primary research.
Section~\ref{sec:conclusion} concludes the paper with an outlook.

\section{A Brief History of Program Comprehension Theory}
\label{sec:history}

We travel a bit in time, more precisely to the 1980s, a decade that introduced almost all the relevant concepts that are still shaping our current picture of code comprehension.
At that time, the term \emph{program comprehension} was mainly used, which today covers a broader range of research topics, of which code comprehension is one~\cite{Wyrich:2022:40Years}.
Program comprehension research of that time is in many aspects comparable to contemporary research. 
However, the period was particularly characterized by the development of models and theories to explain behavioral processes of developers.

We start in 1978, when Ruven Brooks~\cite{Brooks:1978:BehavProgramComp} lays the first milestone of our journey with an essay about a behavioral theory of program comprehension.
He assumes that developers switch between \enquote*{knowledge domains} in program understanding, which, roughly summarized, represent different levels of abstraction of reality.
According to Brooks, a developer has understood a program when not only information about objects and their relationships within one abstraction level are known, but also their relationship in a nearby abstraction.
Brooks' most influential idea, however, was that the process of understanding is characterized by the \enquote{successive refinement of \textbf{hypotheses} about the program's operation}~\cite{Brooks:1978:BehavProgramComp}.
If a developer takes a particularly long time to understand a certain code snippet, this can be explained by the difficulty of finding correct hypotheses at that moment.
Brooks further already speculated at this time that there were elements in the program, such as comments or variable names, that represented \emph{cues} in understanding the program.

Five years later, in 1983, \citet{Brooks:1983:TowardsTheory} elaborated his thoughts and the model that we know today as the \enquote*{\textbf{top-down model}} of program comprehension emerged: Hypotheses about knowledge domains are generated, tested, and refined until their relationship to the code becomes apparent.
A hypothesis describes, for example, the basic function of a component.
The first hypothesis is generated as soon as the developer receives the first information about the program; how appropriate this hypothesis is depends on the skill level and the experience of the developer with the problem domain.
The verification and the establishment of alternative hypotheses are supported by certain indicators in the code.
Brooks calls these indicators \enquote*{\textbf{beacons}} here for the first time.

Between these two publications by Brooks, in 1979 \citet{Shneiderman:1979:Syntactic} came forward with their model for programmer behavior.
In their work, they did not limit themselves exclusively to program comprehension.
Five programming tasks were studied, one of which constitutes comprehension.
Nevertheless, we can also call this work a seminal work because it resulted in what we know today as the \enquote*{\textbf{bottom-up model}} of program comprehension: \enquote{Instead of absorbing the program on a character-by-character basis, programmers recognize the function of groups of statements, and then piece together these \textbf{chunks} to form even larger chunks until the entire program is comprehended}.
\citet{Shneiderman:1979:Syntactic} further emphasized the role of syntactic knowledge (programming language dependent) and semantic knowledge (general programming concepts) in a developer's long-term \textbf{memory}, as well as the role of short-term and working memory for the construction of a \enquote{multileveled internal semantic structure to represent the program [\ldots] The central contention is that programmers develop an internal semantic structure to represent the syntax of the program, but that they do not memorize or comprehend the program in a line-by-line form based on the syntax}~\cite{Shneiderman:1979:Syntactic}.

Both \citet{Brooks:1983:TowardsTheory} and \citet{Shneiderman:1979:Syntactic} have supported their presented concepts with initial experiments, and it did not take long for independent studies to follow that put the models to the test:
\citet{Basili:1982:Understanding} observed themselves comprehending and judging the correctness of a program, providing early anecdotal evidence that bottom-up comprehension practically takes place.
\citet{Adelson:1981:ProblemSolving} found experts to have larger recall chunks than novices, and that expert chunks contain more semantically rather than syntactically related information.
\citet{Wiedenbeck:1986:Beacons} conducted two experiments on the differing influence of beacons on experienced and inexperienced developers, and found that supposedly useful indicators in the code primarily help experienced developers understand it.
Wiedenbeck discusses the role of beacons and considers them to be in line with the bottom-up comprehension strategy, since beacons can represent \enquote{the link between the top-down hypothesis formation stage and the data of the program text}.
Moreover, what Shneiderman calls a chunk could also represent a beacon, provided that the chunk represents a stereotypical part of the program~\cite{Wiedenbeck:1986:Beacons}.
\citet{Soloway:1984:Empirical}, at the same time as \citet{Brooks:1983:TowardsTheory}, investigated top-down comprehension assumptions and provided evidence that \emph{programming plans} and \emph{rules of programming discourse} would help experienced developers with program comprehension.
Programming plans represent stereotypical program fragments, similar to a kind of beacon.
Rules of programming discourse represent intuitive expectations of developers, e.g. that a function name matches the content of the function~\cite{Soloway:1984:Empirical}.
\citet{Rist:1986:Plans} refinded \citeauthor{Soloway:1984:Empirical}'s plan idea and provided additional experimental evidence for its positive influence on program comprehension.

\subsection*{From Comprehension Strategies to Mental Models}

In 1987, the focus of research seemed to shift minimally from cognitive comprehension strategies to the nature of mental representations of source code (mental models).

\citet{Pennington:1987:Stimulus} shows with an example of a code snippet that a program can be abstracted in at least four different ways: (1) by the goals of the program, (2) as a data flow abstraction, (3) as a control flow abstraction, and (4) of \enquote{conditionalized action}, i.e. under certain conditions the program performs actions and enters another state.
Pennington calls the control-flow representation \enquote*{\textbf{program model}} and the combination of data-flow and goal hierarchy representation \enquote*{\textbf{situation model}}, drawing on theories from earlier work in the field of text comprehension~\cite{VanDijk:1983:Strategies}.
In her first study, Pennington shows that procedural (control flow) rather than functional units (goal hierarchy) \enquote{form the basis of expert programmer's mental representations}, which implies that even experts initially apply bottom-up comprehension strategies~\cite{Pennington:1987:Stimulus}.
Her second study addresses the limitation of small code snippets.
Once again, an understanding of program control flow and procedures preceded an understanding of program functions.

\citet{Letovsky:1987:Cognitive} proposes a cognitive model of the \enquote*{knowledge-based program understander} who is made of three components, namely a knowledge base, a mental model of their current and evolving understanding, and an assimilation process interacting with the stimulus materials.
His position is that different types of knowledge play the central role in the comprehension process and that the understander \enquote{is best viewed as an opportunistic processor capable of exploiting both bottom-up and top-down cues as they become available}.
Letovsky elaborates on some details and examines his assumptions using video recordings of six professional developers who were asked to add a new feature to a project while thinking out loud.
What is interesting for us at a later point of this paper is what Letovsky believes a mental model should contain: a description of the goals of the program (specification), a description of actions and data structures in the program (implementation), and an explanation of the links between goal and implementation (annotation)~\cite{Letovsky:1987:Cognitive}.

Finally, \citet{Littman:1987:MentalModels} distinguish between static knowledge (what the program does functionally) and causal knowledge (about interactions of functional components during execution).
Both types of knowledge are required for comprehension, which is why they call a mental model that contains both types of knowledge a \enquote*{strong mental model}.
In contrast, a mental model of the program that only contains static knowledge is called a \enquote*{weak mental model}.
They found some developers to use a \emph{systematic} comprehension strategy, i.e. \enquote{the programmer performs extensive symbolic execution of the data flow paths between subroutines}, and some to use an \emph{as-needed} strategy, meaning that they attempt to only focus on those parts of a program relevant for the experimental task, which resulted in either a weak or a strong mental model.
As a consequence, developers following the systematic strategy were more successful at the requested modification task~\cite{Littman:1987:MentalModels}.
\citet{Koenemann:1991:ExpertStrategies} provided further evidence a few years later that developers follow an \enquote*{as-needed} strategy when they need to understand or modify a program, and that \emph{need} is primarily based on the developer's goals.
They further conclude that program comprehension is mainly a top-down process and bottom-up is used \enquote{in cases of missing or failing hypotheses and locally for directly relevant code units}~\cite{Koenemann:1991:ExpertStrategies}.

\subsection*{The 1990s and the Integrated Model}

We see that there has been neither a lack of concepts that contribute to understanding program comprehension nor a lack of empirical investigations of them. 
In the 1990s, we also find good examples of work that derives implications for practice based on the basic research on program comprehension described so far.
For example, \citet{Storey:1999:CognitiveExploration} address the design of software exploration tools to support developers and use theory to explain what design principles the tools would need to follow to facilitate the construction of a mental model for program understanding.

\begin{sloppypar}
What was missing, however, was some kind of metamodel that coherently integrated the various concepts of the 1980s.
\citet{Vonmayrhauser:1995:PCduring} took on this task and in 1995 set another milestone in our history of program comprehension theory with their \enquote*{integrated metamodel} of comprehension strategies.
\end{sloppypar}

\begin{figure}[ht]
    \centering
    \includegraphics[width=\columnwidth]{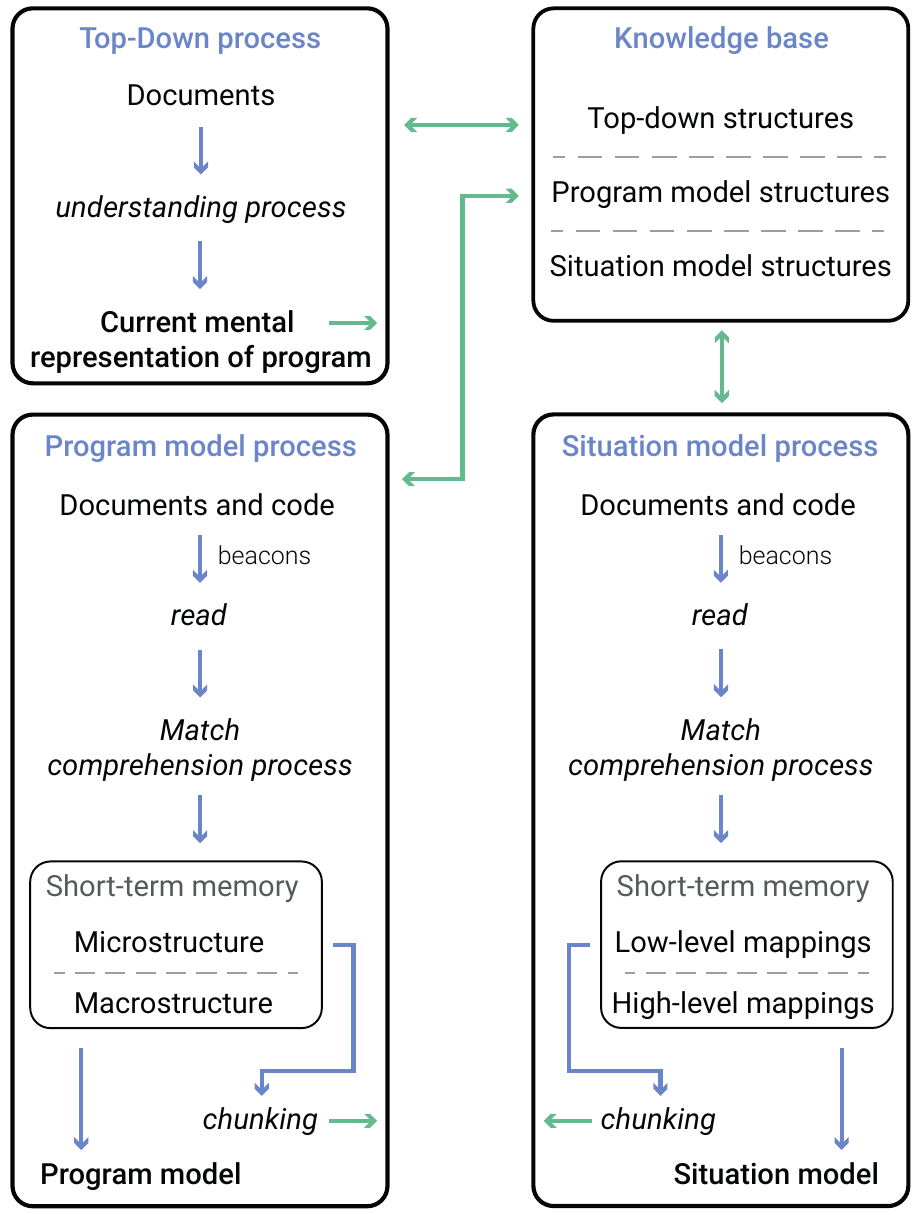}
    \caption[The \enquote*{integrated metamodel} by \citeauthor{Vonmayrhauser:1995:PCduring}]{The \enquote{integrated metamodel} by \citeauthor{Vonmayrhauser:1995:PCduring}. Figure based on Figure 6 in~\cite{Vonmayrhauser:1995:PCduring}.}
    \label{fig:integratedModel}
\end{figure}

Figure~\ref{fig:integratedModel} depicts the metamodel.
It is a re-creation of the graphical illustration of von Mayrhauser and Vans, where we have simplified the knowledge base in its level of detail to make the illustration more concise and understandable.
Apart from that, all components are included that are also contained in the original model: The top-down model based on \citet{Soloway:1984:Empirical}, the program model and situation model based on \citet{Pennington:1987:Stimulus}, and the knowledge base, as the node that contains the information to build and switch between the respective cognitive process models.
The integrated metamodel also incorporates elements from the work of \citet{Brooks:1983:TowardsTheory}, \citet{Letovsky:1987:Cognitive}, and \citet{Shneiderman:1979:Syntactic}, so that we find again, e.g., the concepts of beacons, memory, chunking, and various kinds of knowledge.
All five incorporated models further describe a matching process between the developer's knowledge and the object to be understood (\enquote*{match comprehension process}), for example through the systematic verification of hypotheses according to \citet{Brooks:1983:TowardsTheory} or via a more opportunistic approach according to \citet{Letovsky:1987:Cognitive}.

Back then, von Mayrhauser and Vans considered the development of program comprehension models to be \enquote{in the theory-building phase} and theories on large-scale program comprehension to be \enquote{in their infancy}~\cite{Vonmayrhauser:1995:PCduring}.
They could not have known that the end of the 1990s also marked the end of two decades of research that produced the most influential models of program comprehension to date.

Before we turn to the progress towards theory building in the 2000s, we would like to highlight an aspect that was increasingly made explicit in papers of the 1990s and seems still a consensus today~\cite{Walid:2014:ComprehensionOfPC}: program comprehension is usually not an end in itself, but a (necessary) means to accomplish a maintenance task successfully.
We might do well, therefore, to also develop models that are specific to particular development activities and incorporate program comprehension as one aspect of such activity.

\citet{Gilmore:1991:ModelsDebugging} presents an example of such a model that connects the debugging and comprehension processes. 
Gilmore argues that the view that developers use a mental model created during the comprehension process in a separate step for debugging is not accurate.
Instead, he argues that these are two distinct processes, of which comprehension is part of a larger debugging loop: \enquote{The fact that, in both cases, bugs were more easily found when they were deeper in the structures contradicts the traditional interpretation of propositional analyses of comprehension. Both groups of programmers were more likely to find the deeper bugs, suggesting that they were being detected during an ongoing comprehension process, rather than by comparison with a comprehension state}~\cite{Gilmore:1991:ModelsDebugging}.
Gilmore's model not only embeds the comprehension process in a higher-level debugging activity, it also models the interesting aspect of \emph{problem comprehension} and places mismatch detection between the mental representation of the program and the mental representation of the problem at the center of debugging.

\subsection*{The 2000s: Model Refinements, Model Extensions, and\ldots Teaching}

We enter the decade in which the International Workshop on Program Comprehension transitioned into a full conference.
In the context of program comprehension strategies and theory, 
some existing models were refined or extended.

\citet{Crosby:2002:RolesBeacons} conducted further research on the role of beacons.
Among other methods, they used eye-tracking to see what developers fixated on and for how long.
They found that there are at least two types of beacons.
One is single lines of code, which contain mnemonic cues about functionality, and the other is more complex beacons, which contain cues about main goals of the program.
Also, experts seem to rely on beacons, while novices do not.

At the same time, \citet{Rajlich:2002:RoleOfConcepts} discuss the role of \enquote*{concepts} for understanding and learning about unknown and large programs.
They define concepts as \enquote{units of human knowledge that can be processed by the human mind (short-term memory) in one instance}.
Examples of concepts can be single features of the software, which can be found as such in the code. There may be an overlap here with the concept of complex beacons described by \citet{Crosby:2002:RolesBeacons} and there is definitely some relationship to the idea of chunking parts of the code into larger segments~\cite{Shneiderman:1979:Syntactic}.
In any case, \citet{Rajlich:2002:RoleOfConcepts} consider domain concept knowledge essential for program comprehension.

Building on the works of \citet{Brooks:1983:TowardsTheory} and \citet{Soloway:1984:Empirical}, \citet{OBrien:2004:BottomUp} refine the idea of top-down comprehension into expectation-based and inference-based comprehension, where the difference can be roughly summarized as whether beacons~\cite{Rist:1986:Plans,Wiedenbeck:1986:Beacons} in the code confirm or trigger a developer's (pre-generated) hypotheses.
In an experiment with eight industrial developers, they were able to show empirically that for expert programmers both strategies play a role together with the bottom-up comprehension strategy.
The preferred approach depends on the familiarity with the application domain~\cite{OBrien:2004:BottomUp}.

\citet{Gueheneuc:2009:Theory} extends existing theories on program comprehension by adding vision science, i.e. the \emph{recognition} of items during program comprehension.
The work brings an interesting new dimension into play: modality.
In the code comprehension experiment landscape, modality has so far been somewhat neglected, perhaps because a homogeneous picture of the code-\emph{reading} developer prevails in the minds of the researchers, perhaps also because modality is assumed to have little influence on the comprehension of information in general~\cite{Rogowsky:2016:modality}.

In addition to these theoretical extensions and refinements, the theory of program comprehension strategies of the 80s and 90s also increasingly found its way into teaching.
A workshop working session on teaching program comprehension was organized with the goal of establishing \enquote{a repository of teaching resources, to identify the lessons learned and future directions}~\cite{Deursen:2003:Teaching}.
Then, for example, \citet{Exton:2002:Constructivism} reviews existing program comprehension strategies and its relation to the constructivist learning theory, and \citet{Schulte:2008:BlockModel} proposes a model of core aspects of understanding a program text that should guide instructors in designing teaching lessons.
The interested reader will find a good introduction to how program comprehension models can enrich the computer science educational perspective in \citet{Schulte:2010:IntroductionEducators}.

\subsection*{The 2010s: What Happens in Practice?}

A few refinements of theoretical program comprehension models also entered the scientific literature in the 2010s~\cite{Belmonte:2014:ThreeLayerModel,Nosal:2015:FourLayerModel,Shargabi:2015:LevelsOfAbstraction}, but the decade was marked in particular by studies that examined program comprehension in practice.

\citet{Walid:2014:ComprehensionOfPC} conducted a mixed methods study in which they observed and interviewed 28 software professionals and surveyed 1477 others about their comprehension strategies, the nature of the knowledge relevant to understanding, and the use of comprehension tools.
The work is a goldmine of qualitative and quantitative findings.
Some of the key insights are that developers use a recurring, structured comprehension strategy depending on the task context, the starting point for comprehension depends on experience, hypotheses are indeed generated and tested as described by the theoretical models, and that developers sometimes take notes to reflect on their mental model.
Further, source code is more trusted than documentation, and naming conventions and common architecture simplify program comprehension.
\citet{Walid:2014:ComprehensionOfPC} also found that dedicated program comprehension tools are not used.
Developers would instead prefer to use basic tools such as compilers and text editors for program comprehension.

\citet{Minelli:2015:LastSummer} suggest that about 70 percent of a developer's time is spent on program comprehension. 
They quantitatively examined an IDE interaction dataset of 740 development sessions from 18 developers (mouse and keyboard events together with contextual information).
A development activity was defined as a sequence of mouse and keyboard sprees.
Development activities were then categorized into understanding, navigation, editing, and UI interactions, where understanding consisted of basic understanding, inspection, and mouse drifting (i.e. mouse-supported reading).
The results show that by far the most time is spent on understanding (70\%), followed at a large distance by UI interactions (14\%), editing (5\%), and navigation (4\%), with the remaining 7--8\% being time spent outside the IDE.
The authors highlight the importance of not getting interrupted during developmental activity, as it is largely characterized by mental processes.

On a side note, both \citet{Walid:2014:ComprehensionOfPC} and \citet{Minelli:2015:LastSummer}, arrive at a similar picture of program comprehension as we discussed earlier, namely as an embedded activity.
They conclude that \enquote{program comprehension is considered a necessary step to accomplishing different maintenance tasks rather than a goal by itself}~\cite{Walid:2014:ComprehensionOfPC} and \enquote{that base understanding is prevalent inside activities, that is, inside conceptually related sequences of keyboard or mouse sprees. In other words, the process of program understanding is not really an activity per-se, but it is interleaved with other activities like editing}~\cite{Minelli:2015:LastSummer}.

Just three years after \citeauthor{Minelli:2015:LastSummer}'s research, the key message that program comprehension accounts for the majority of developmental activity was confirmed by \citet{Xia:2018:Measuring}.
They conducted a large-scale field study with 78 software professionals from two IT companies in China and found that developers spend about 58 percent of their time on program comprehension activities.
Since, according to the authors, program comprehension does not only take place in the IDE, \citet{Xia:2018:Measuring} tracked the activities of developers across different applications.
They were able to show that only about 20\% of program comprehension takes place in the IDE, 27\% in web browsers and about 10\% in document editors.
There is frequent switching between these applications.

\subsection*{\textbf{The 2020s: Dawn of the Neuro Age}}

Up to this point, we have observed that much has been done in terms of fundamental theories around program comprehension, especially in the 1980s and 1990s, but these efforts have since given way to other efforts, such as integrating the theories into teaching or comparing them to practitioners' experiences.
Toward the middle and end of the 2010s, a new trend emerged that we predict will dominate the 2020s and may revive basic research: the application of psycho-physiological measurement methods in the context of code comprehension.

While its realization brings some new challenges, the idea is simple to explain: Researchers use eye tracking, functional magnetic resonance imaging (fMRI), functional near-infrared spectroscopy (fNIRS), electroencephalography (EEG), or combinations of these devices to gain insights into the cognitive processes associated with software engineering tasks (neural correlates) to better understand and support such software engineering tasks~\cite{Vieira:2021:Psychophysio,Sharafi:2021:ObjectiveMeasure,Siegmund:2020:Studying}.
A recent systematic mapping study in software engineering found that most papers that make use of psycho-physiological data explore program comprehension and debugging strategies~\cite{Vieira:2021:Psychophysio}.

Except for eye tracking, which was already used in program comprehension studies two decades ago (e.g.~\cite{Crosby:2002:RolesBeacons}), the number of studies using, for example, fMRI and fNIRS to study program comprehension is now also increasing.
Some position papers promisingly outline the possibilities of gaining more objective insights into program comprehension with neuroimaging devices~\cite{Fakhoury:2018:ObjectiveMeasures} or even an entirely new, neurocognitive perspective on program comprehension~\cite{Peitek:2018:NeuroCogPerspective,Siegmund:2020:Studying}.
However, this development of the research field is still at a stage where some researchers would agree that the achieved reliability of the related research setups is probably the greater achievement to date than initial findings on brain activities in program understanding~\cite{Siegmund:2020:Studying}.
Nonetheless, there are initial findings from psycho-physiological research that strengthen and expand our understanding of program comprehension.

\citet{Siegmund:2014:UnderstandingUnderstanding} conducted the first fMRI study in which participants had to understand code while positioned in a scanner. Measured blood oxygenation levels can be used to determine brain regions that are active at the time of a code comprehension task. To identify the regions that are responsible for comprehension in isolation, participants also had to perform a contrast task of finding syntax errors. The data show increased activity in several Brodmann areas during code comprehension. The authors identified five of these as relevant for code comprehension (BAs 6, 21, 40, 44, 47), since it is known from previous studies that these areas are related to working memory, attention, and language processing. The findings form a basis for the development of a cognitive model of bottom-up code comprehension that could enrich the previously discussed theory~\cite{Siegmund:2014:UnderstandingUnderstanding}.

Further studies followed that at least partially confirmed these initial findings and provided some additional insights~\cite{Siegmund:2017:NeuralEfficiency,Peitek:2018:Simultaneous,Floyd:2017:DecodingFMRI,Lee:2016:ComparingEEG}. For example, another study by \citet{Siegmund:2017:NeuralEfficiency} showed that the presence of semantic beacons in code leads to lower activity in the relevant brain regions during comprehension, and \citet{Floyd:2017:DecodingFMRI} found that developer expertise affects how accurately a classifier can distinguish code and prose review based on brain activity.
Also, worth mentioning is \citeauthor{Peitek:2018:Simultaneous}'s study~\cite{Peitek:2018:Simultaneous}, in which, for the first time, fMRI and eye tracking were tested in combination to investigate code comprehension. Such a combination has the advantage that fast cognitive subprocesses can be captured due to a higher temporal resolution of the eye tracker. This allowed them to show, for example, that fixations on a beacon leads to semantic recall~\cite{Peitek:2018:Simultaneous}.
We refer the interested reader to \citeauthor{Vieira:2021:Psychophysio}'s mapping study~\cite{Vieira:2021:Psychophysio}, \citeauthor{Siegmund:2020:Studying}'s viewpoint paper~\cite{Siegmund:2020:Studying}, and to the introduction of the paper by \citet{Sharafi:2021:ObjectiveMeasure} for further pointers to recent psycho-physiological studies in software engineering.

\section{Defining and Conceptualizing Source Code Comprehension}
\label{sec:model}

Much has been done to understand \emph{how} developers approach what has been termed program or code comprehension right from the creation of the construct.
We have learned that the construct can be operationalized in several ways, and that various external factors as well as the comprehension strategy itself can influence the comprehension success measures employed.
However, what literature has yet to provide us is a definition of \emph{what} code comprehension can be on a conceptual level.
We will therefore define source code comprehension in this section and then design a conceptual model of code comprehension experiments.

\subsection{A Definition of Source Code Comprehension}

The landscape of prominent definitions and approaches to such definitions for code comprehension is quite limited to the early days of the research field.
For \citet{Shneiderman:1977:Measuring}, for example, program comprehension is \enquote{the recognition of the overall function of the program, an understanding of intermediate level processes including program organization and comprehension of the function of each statement in a program}.
Similarly, \citet{Letovsky:1986:DelocalizedPlans} state that \enquote{the goal of program understanding is to recover the intentions behind the code} and \citet{Pennington:1987:Stimulus} elaborates that \enquote{comprehension involves the assignment of meaning to a particular program, an accompplishment that requires the extensive application of specialized knowledge}.

Further, \citet{Gilmore:1991:ModelsDebugging} made a valuable theoretical contribution by explicitly distinguishing between the comprehension process and the state of comprehension: \enquote{The former involves the mobilisation of cognitive resources and processes in some particular configuration, with the goal of constructing some mental representation of the program code. It is this mental representation of the code which is the comprehension state}.

Our definition builds on this foundation.
In particular, the recovery of intentions (\citet{Letovsky:1986:DelocalizedPlans}) and meaning (\citet{Pennington:1987:Stimulus}) behind the code, and Gilmore's distinction between comprehension process and comprehension state are reflected in the definition below.
Moreover, our definition deliberately uses the term \emph{source code} instead of \emph{program} to explicitly distinguish code comprehension from comprehension of other objects in the context of the diversified research field of program comprehension that exists today.
Accordingly, we consider the introduction of a new definition of the construct to be justified by the way in which it unifies historically evolved views in a contemporary manner.\\

\noindent\textbf{Definition 3.1 (Source Code Comprehension)}\\
\textit{Source code comprehension describes a person's intentional act and degree of accomplishment in inferring the meaning of source code.}\\

Accompanying this definition are some propositions that explain the construct in more detail:
\begin{enumerate}
    \item The \emph{degree of accomplishment} represents a spectrum. Colloquially, we tend to speak of someone having understood or not understood something, thus treating the construct as a binary variable. Researchers are free to do the same in their operationalization of the construct, but in doing so, they would miss out on potential in the finer-grained analysis of performance differences of their study participants, and this should be justified argumentatively.
    \item The \emph{degree of accomplishment} can be expressed by contrasting a developer's mental model with external reality in terms of the developer's goals. Such a comparison suggests intuitively that one measures how accurately and completely the mental model reflects reality. However, it should be kept in mind that, depending on the goal of the developer, it is not always favorable to aim for a complete mental model of the code. We learned in Section~\ref{sec:history} that the formation of mental models depends on both the comprehension strategy and the comprehension task. For example, the task of understanding source code in detail and the task of finding a bug in the code will lead to different mental models of the code (i.e., what \citet{Koenemann:1991:ExpertStrategies} described as \enquote*{as-needed} strategy). When designing the concrete way to measure the success of code comprehension, this aspect must be considered.
    \item \emph{Inferring the meaning of source code} can be discussed and assessed on at least three dimensions: what the code does (functionality), what it should do (specification), and what the intention of the code author was (context). These aspects can be found in similar forms in various program comprehension models and are surveyed to varying degrees in contemporary code comprehension studies. Again, there is no single approach here: The concrete task design depends on the reasoning of the researcher regarding the goal scenario of the developer.
    \item There are a variety of ways to observe and measure source code comprehension. The definition applies to behavioral and neuro-physiological correlates, as well as both assessments in the form of administered comprehension tests and self-assessments of the subject.
\end{enumerate}

Source code comprehension is to be distinguished from source code comprehensibility.
\emph{Code comprehension} is the subject of the definition and describes the act and degree of accomplishment in understanding code.
\emph{Code comprehensibility}, on the other hand, describes a quality attribute of the code. 
Empirical studies that intend to provide insights into the influence of a treatment on the comprehensibility of code usually first measure code comprehension, i.e., how well a participant understood source code under certain conditions.
The data collected can then be used to draw conclusions about the comprehensibility of the code.
Further, we consider \emph{comprehensibility} and \emph{understandability} as synonyms, i.e., choosing one over the other is \enquote{purely a matter of linguistic variation}~\cite{Kintsch:1998:Comprehension}.

A related construct is code readability, a property of code that indicates the ease with which the code can be read. Influences on readability can be, for example, the choice of simple variable names or familiar control structures. 
However, readable code does not necessarily have to be understood by the reader.
In both code and text comprehension, there is evidence that comprehensibility and readability do not necessarily correlate~\cite{Borstler:2016:RoleMethodChains,Smith:1992:ReadabilityUnderstandability}.

\subsection{A Conceptual Model for Code Comprehension Experiments}

\epigraph{\textit{The goal was always the same: to develop models and tools to help developers with program understanding during program maintenance. However, few authors targeted the more fundamental question: \enquote{what is program understanding?}}}{\citet{Harth:2017:ProgramUnderstandingModels}}

There is a sentiment in \citeauthor{Harth:2017:ProgramUnderstandingModels}'s quote that we can relate to a certain extent.
Especially in contemporary code comprehension literature, there are hardly any explanations of what the authors mean by code understanding.
To be fair, though, the question of what \enquote*{understanding} means conceptually is not a simple one.
Often, the answer also depends on the specific context, for example, whether it is a question of implementing an AI with the capability of understanding user needs or whether it is a question of explaining neuroscientifically how information processing takes place at different levels of the nervous system\footnote{Kevin Mitchell, in a blog post about whether an AI could cook meth, conveys the potential diversity of views well: \url{http://www.wiringthebrain.com/2019/02/understanding-understanding-could-ai.html}}.

In our context in which developers need to understand source code, the above Definition 3.1 and associated propositions provide a starting point.
We now take this one step further and describe on a conceptual level what code comprehension can mean in the context of empirical studies.
Thereby, we restrict ourselves to empirical studies with an experimental character\footnote{We are aware that the term experiment is used differently~\cite{Stol:2018:ABC} or even incorrectly~\cite{Ayala:2022:UseMisuseExperiment} in software engineering research, and that there are different types of experiments.
The experiments we focus on here are closest to what \citeauthor{Stol:2018:ABC} classify as \enquote*{Laboraty Experiment}, the purpose of which is \enquote{to study with a high degree of precision relationships between variables, or comparisons between techniques; may allow establishment of causality between variables}~\cite{Stol:2018:ABC}.
These can then be further subdivided into, for example, randomized controlled experiments and quasi experiments.} to ensure a certain comparability of the study designs.
And we remain at a level of abstraction that allows us to map the majority of existing code comprehension experiments to the proposed conceptual model.

Our conceptual model for the design of code comprehension experiments is depicted in Figure~\ref{fig:comprehensionExperimentFramework}.
The model consists of three horizontal lanes: mental model, mental state, and experimental variables.
These play a role at two points in time $t$ and $t+x$ with $x > 0$.

\subsubsection*{\textbf{The Lanes of the Model}}

Let us start with the top two lanes and their interaction.
At each point in time, a developer is in a mental state $S$, which is backed up by a mental model $M$ with a (partial) representation of the source code to be understood.
The beginning of an experiment may represent time $t$; at this time, a developer's mental model of a particular code snippet (depending on prior knowledge of the code) may not be too extensive.
Later in the experiment ($t+x$), after completing a code comprehension task, the developer is in a new mental state and accordingly has a changed mental model of the code.

A lot can happen between these two points in time.
Developers generate hypotheses about how the code works, reject or confirm them, and understand the code better over time.
This process generates any number of mental states and changes in the mental model between the two explicitly depicted mental states.
The two mental states $S_{t}$ and $S_{t+x}$ are particularly crucial for designing code comprehension experiments because they represent the initial state and the state in which a researcher tries to access the internal mental model via externally observable proxies to infer how well a developer has understood the source code presented (this may happen several times during the experiment).

\begin{figure}[t!]
    \centering
    \includegraphics[width=\columnwidth]{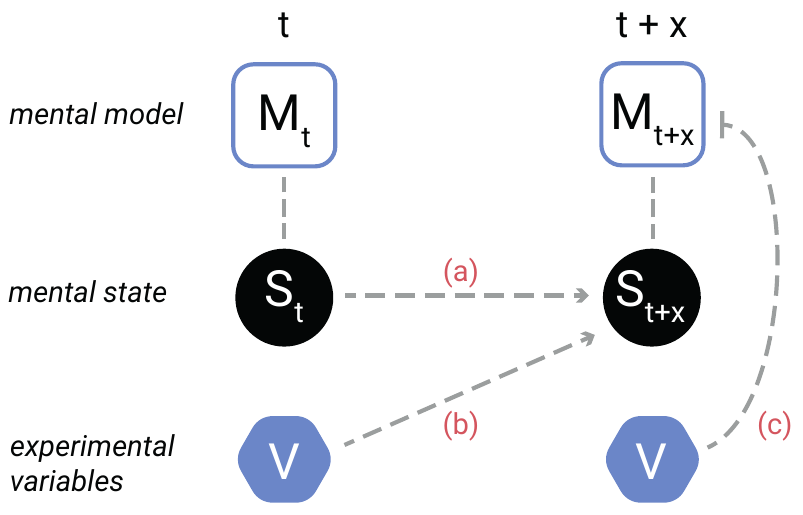}
    \caption{Conceptual model for the design of code comprehension experiments in which the focus is on measuring a participant's eventual comprehension of a code snippet based on their formed mental model \textcolor{colorB}{(c)}.}
    \label{fig:comprehensionExperimentFramework}
\end{figure}

The third lane, experimental variables, includes all experimental design decisions (e.g. the comprehension task), contextual factors (e.g. time pressure, cognitive biases), individual characteristics of the developer (e.g. programming experience), and other variables that can influence the mental state of a developer.
In Figure~\ref{fig:comprehensionExperimentFramework} we see that the mental state $S_{t+x}$ is influenced by both the previous mental state $S_{t}$ and a set of experimental variables $V$.
Part of the experimental variables, namely the statistical tests and comprehension measures, operationalize in which way the mental model $M_{t+x}$ is accessed at the time $t+x$.

The term \enquote*{mental model} is used in this context as it was introduced in the 1990s in the context of program comprehension strategies, i.e. as \enquote{a maintainer's mental representation of the program to be understood}~\cite{Storey:1999:CognitiveExploration} and \enquote{an internal, working representation of the software under consideration}~\cite{Vonmayrhauser:1995:PCduring}.
According to a recent meta-study on mental representations during program comprehension, this description captures the essence on which the community could agree~\cite{Heinonen:2022:SynthesizingMentalModels}.
At the same time, it should be mentioned that it would be difficult to find controversies at present anyway, since research in the field of mental representations during program comprehension has unfortunately declined considerably~\cite{Bidlake:2020:ReviewRepresentations,Heinonen:2022:SynthesizingMentalModels}.

The meaning of the term \enquote*{mental state} is quite diverse due to different streams in the philosophy of mind.
For example, for followers of physicalism mental states are equal to brain states (followers of this theory are in the majority today), for followers of dualism theory mental and physical are two different things~\cite{Jaworski:2011:Mind,Kind:2020:Mind}. 
The topic is as exciting as it is complex because between and alongside these positions there are many variants, all of which sound plausible at first, but over time all of which have at least been confronted with thought experiments, and from some people's point of view also refuted.
Further elaboration is beyond the scope of this paper, so we handle the matter pragmatically as follows.
Understanding is directly related to the mind, and a mental state can be considered as the mind of a person at a particular time.
Therefore, we would like to reflect this connection in the conceptual model, in that mental states represent the bridge between the environment (experimental variables) and the mental model a developer has about the code.
However, the conceptual model allows for different views on the nature of mental states, so that in the first place, we only have to agree on the idea that at any point in time such a state exists, which will transition into another state (similar to a state machine).
This perspective allows us to model a progression in the comprehension process of a developer.

\subsubsection*{\textbf{The Links of Interest}}

With this, we have a rough understanding of the elements of the conceptual model for the design of code comprehension experiments.
We will now go into more detail on the three depicted links (a), (b), and (c) in Figure~\ref{fig:comprehensionExperimentFramework}.

Link (a) represents the transition from one mental state to another and thus constitutes the \emph{cognitive model}, i.e., \enquote{the cognitive processes and information structures used to form the mental model}~\cite{Storey:1999:CognitiveExploration}.
There is research on this, as we saw in Section~\ref{sec:history}, but it is far from sufficient to adequately characterize the processes and information structures involved in code comprehension.

Link (b) is about the (assumed) influence of experimental variables on the code comprehension process and accomplishment.
Any experiment that investigates the influence of a variable on a participant's eventual code comprehension is interested in this link (b).
It is important to understand that each such experiment creates its own implicit or explicit causal model.
While link (b) is drawn as a single line with an arrowhead at the level of abstraction of our model (Figure~\ref{fig:comprehensionExperimentFramework}), one would see at a finer level of detail that it is actually a complex network of interdependent variables.
Examples of code comprehension studies in which such a causal model has been made explicit can be found in \cite{Wagner:2021:Intelligence,Wyrich:2022:Anchoring}.
Moreover, please note that link (b) can be directed in the opposite direction, from a mental state to a later relevant experimental variable, that is, when a study intends to investigate the consequences of code comprehension on a different construct (e.g., affective states).

Link (c) deals with capturing the mental model, relevant when measuring a person's code snippet understanding.
In most experiments, this happens once per experimental task, for example when comprehension questions are asked to the participant after understanding a code snippet.

\begin{figure}[t!]
    \centering
    \includegraphics[width=\columnwidth]{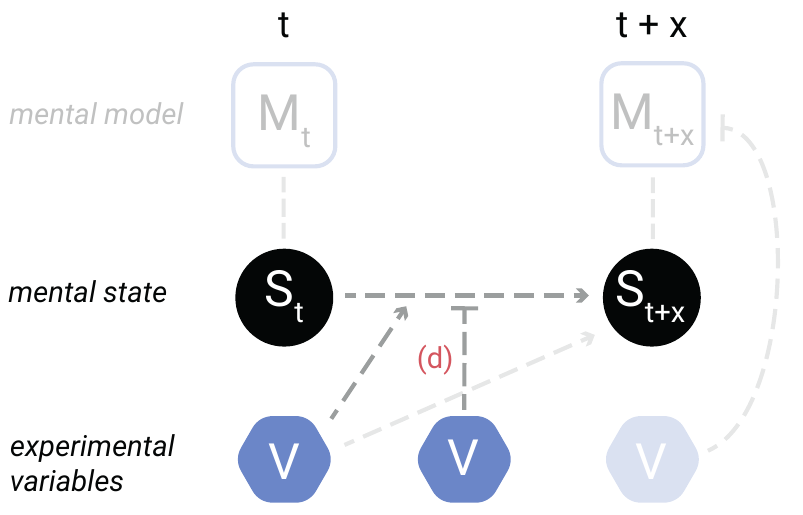}
    \caption{Conceptual model for the design of code comprehension experiments in which the focus is on observing the cognitive processes of comprehension \textcolor{colorB}{(d)}.}
    \label{fig:comprehensionExperimentFramework2}
\end{figure}

\subsubsection*{\textbf{A Variant With Focus on Observation}}

Nowadays, we find more and more empirical code comprehension studies in which the focus is on the investigation of link (a), the cognitive processes involved in code comprehension themselves, for example in the investigation of activated brain regions during code comprehension tasks.
In some of these studies, the assessment of the mental model as depicted in link (c) of Figure~\ref{fig:comprehensionExperimentFramework} becomes secondary to irrelevant.\\
Figure~\ref{fig:comprehensionExperimentFramework2} shows how the conceptual model can account for such studies.
The operationalization of code comprehension is achieved by experimental variables capable of observing the transition between mental states.
Link (d) depicts such an observation.
In Section~\ref{sec:examples} below, we illustrate in concrete terms what this can look like in a primary study on code comprehension.

\begin{figure}[t!]
    \centering
    \includegraphics[width=\columnwidth]{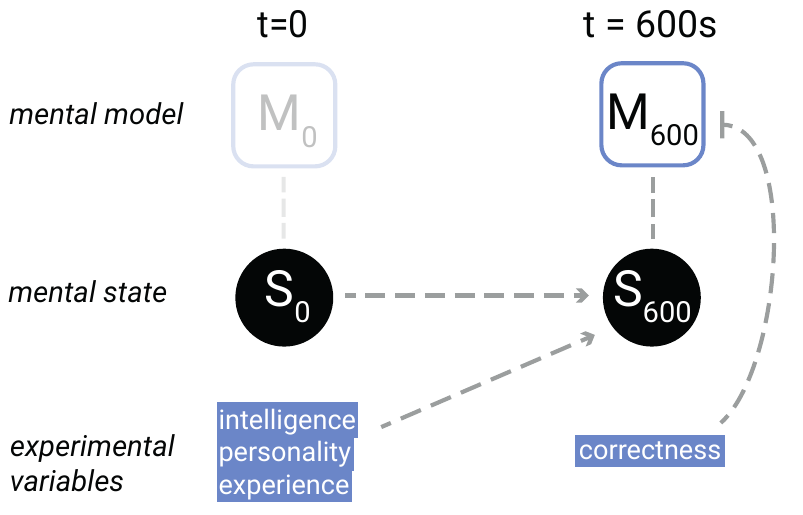}
    \caption{The comprehension part of \citet{Wagner:2021:Intelligence} mapped to the conceptual model for the design of code comprehension experiments.}
    \label{fig:caseIntelligence}
\end{figure}

\section{Application to Case Examples}
\label{sec:examples}

We chose two representative code comprehension experiments for which we argue, based on our presented definition and conceptual model, that these experiments measure the same construct.
We thus take the role of a researcher who conducts a secondary study and has to decide whether certain studies are comparable (we also stress this aspect because the authors of the selected studies may have their own, non-communicated, but divergent view of what they mean by code comprehension).

The first study is by \citet{Wagner:2021:Intelligence} and examines the influence of intelligence facets, personality traits, and programming experience on a developer's accomplishment in comprehending source code.
The comprehension of the study participants is measured by the correctness of the answers to comprehension questions about code snippets.
Here, two dimensions of comprehension are tested: did participants understand what the code is supposed to do (specification), and what does the code actually do for specific input values (functionality). 
Each participant views two code snippets, and has 10 minutes each to complete the comprehension questions.

The authors themselves call the construct of interest either code comprehension or code comprehension performance. 
They define it firstly in the background chapter via the workaround of using \citet{Boehm:1976:Quantitative}'s definition of source code understandability (\enquote{code possesses the characteristic of understandability to the extent that its purpose is clear to the inspector}), and secondly via their operationalization in the methodology section of their paper.
We argue that our proposed definition of source code comprehension applies to the construct that \citeauthor{Wagner:2021:Intelligence} investigated.

Figure~\ref{fig:caseIntelligence} maps their experiment~\cite{Wagner:2021:Intelligence} to our conceptual model.
The time points of interest are t=0, at which time the code comprehension task for a code snippet begins, and t=600s, when the time limit for processing the comprehension task of a snippet is reached.
The authors are interested in how intelligence, personality, and experience influence the mental state and thus the mental model at the end of the task.
At this point, they evaluate the correctness of the given answers.
The authors are not interested in examining the cognitive comprehension process itself, that is, the transition between mental states.

This is where our second case example differs.
\citet{Siegmund:2017:NeuralEfficiency} investigate how the interaction of task (bottom-up comprehension or comprehension based on semantic cues), code layout (pretty-printed or disrupted), and beacons (being present or not) affects neural efficiency and activated brain regions during code comprehension.
The authors measure how blood oxygen levels change while the study participants, lying in an fMRI scanner, have to understand code snippets for different input values and judge whether the snippet correctly implements the same functionality as snippets seen during a training session.
Participants have 30 seconds per snippet to do so.

\begin{figure}[t!]
    \centering
    \includegraphics[width=\columnwidth]{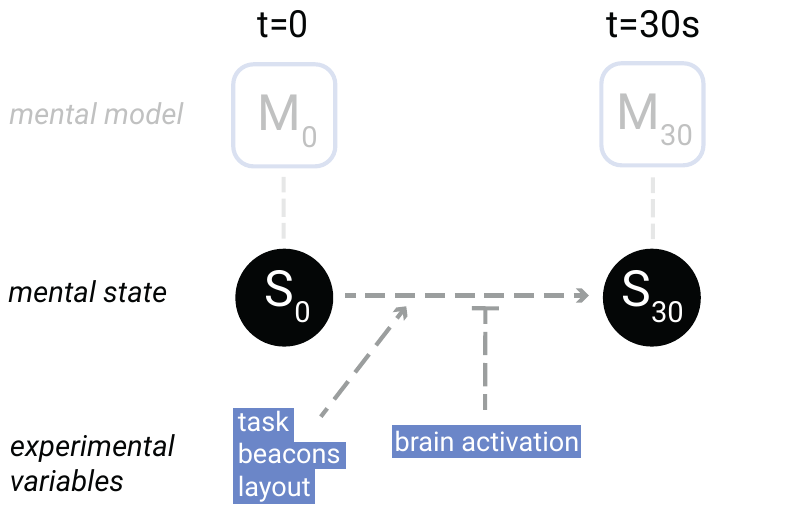}
    \caption{The comprehension part of \citet{Siegmund:2017:NeuralEfficiency} mapped to the conceptual model for the design of code comprehension experiments.}
    \label{fig:caseNeural}
\end{figure}

\citet{Siegmund:2017:NeuralEfficiency} use the term program comprehension and implicitly define the construct under study by distinguishing the bottom-up comprehension strategy and comprehension based on semantic cues (cf. Section~\ref{sec:history}). 
Figure~\ref{fig:caseNeural} maps the comprehension part of their study to our conceptual model.
Of interest to the authors is only the cognitive comprehension process, represented as the transition between mental states at the times of displaying and hiding a code snippet after 30 seconds.
For followers of physicalism, mental states would correspond to brain states in this study.

\subsection{Implications}

We see from the two case examples that the conceptual model covers several of the main research branches of code comprehension research: the more classical behavioral research on the influences on code comprehension (link b and c) and the neuroscientific investigation of code comprehension (link a and d).
Note that both aspects can also occur within a single study, which would result in a model instance comparable to the merging of Figure~\ref{fig:caseIntelligence} and Figure~\ref{fig:caseNeural}.
Additionally, meta-scientific discourse on the operationalization of code comprehension can be anchored within the model (link c and d, respectively).

\subsubsection{Why is our work useful to authors of primary studies?}
As a researcher conducting primary research, it is possible to anchor the study design in the conceptual model. This allows for defining the construct under study and justifying design decisions, such as the operationalization of the construct.
Meaningful discussion of construct validity threats is only possible at all with such a definition~\cite{Sjoberg:2022:Construct}.

We see that in the threats to validity sections of code comprehension experiments, uncertainty about the operationalization of the construct is sometimes expressed quite openly~\cite{Wyrich:2022:40Years}.
While we deliberately do not recommend in this paper how code comprehension should be operationalized, the definition and conceptual model at least help to clarify what code comprehension can be about.
Further, if it becomes apparent during the design process of a study that the design cannot be anchored in the conceptual model presented, this may indicate that one's study may be only comparable to a limited extent with studies that could be anchored in this concrete model.

\subsubsection{Why is our work useful to authors of secondary studies?}
Researchers conducting secondary research usually need to be able to decide whether a particular primary research paper should be included in their dataset.
The ability to map a primary study to the conceptual model can become an explicit inclusion criterion.
This allows inclusion criteria to be evaluated more formally than has been handled to date.
For example, \citet{Wyrich:2022:40Years} note in their systematic mapping study of code comprehension experiments that they had to invest a great deal of time and effort in being able to assess whether a primary study measures code comprehension.
As a consequence, they call on the community to work on a definition of the construct so that researchers of primary studies can use one in their papers.
We respond to their call with our work.

\subsubsection{Why is our work useful for the meta-scientific discourse?}
Our work encourages discussions about whether the prevalent diversity in the design of tasks and measures in code comprehension studies may be in part because different constructs are measured at the conceptual level.
Specifically, one might begin to distinguish such measures that primarily serve as proxies for code comprehensibility, rather than being able to measure a person's understanding of code or a person's comprehension process.
For example, a study that measures the time to find a bug, and merely uses the measured time as a correlate with another construct, does neither consider a developer's resulting mental model of the code nor the transition between mental states.
It is therefore likely that the mere operationalization via time can serve as a proxy for code comprehensibility, but cannot be anchored in the conceptual model presented.

Apart from motivating such discussions, this work contributes in a broader sense to the maturation of the research field.
After devoting nearly fifty years to studying code comprehension, we have arrived at a definition for the construct of interest.

\subsection{Limitations}

The proposed conceptual model is limited in two aspects.
First, for the conceptual model presented in this paper, we focused on experiments, although definition 3.1 applies regardless of the chosen study methodology.
Observing participants in a more qualitative way, for example in an ethnographic study of code comprehension strategies, can only be anchored in the conceptual model to a limited extent, since the model is mostly reduced to the cognitive comprehension aspect, and the outward conduct of a person is not depicted in this process.
Extensions and modifications of the model are conceivable and desired.
We motivated in the introduction that there is a need for additional explicitly formulated perspectives on what code comprehension can be.

Second, our definition of code comprehension is one that focuses on humans. 
An emerging research theme within the code comprehension research community involves the exploration of large language models and AI agents that can explain code to the developer and thus at least mimic code comprehension.
However, a discussion of when a machine understands source code is left for future work.

\section{Conclusion}
\label{sec:conclusion}

In the introduction, we presented the fictional story of Favian, a visitor to an art museum who unexpectedly encountered source code displayed on a canvas hanging on the gallery wall.
Despite lacking any prior knowledge of code, Favian eventually arrived at a point where he claimed to \enquote{understand} what he saw after contemplating it for some time.
If Favian did indeed try to understand the meaning of the source code depicted, then his approach and accomplishment in code comprehension is consistent with our definition of source code comprehension.
We could then use the proposed conceptual model to check how well Favian actually understood the code.
Alternatively, if Favian interpreted the artwork and the artist's intentions instead, it would represent a departure from our definition, thereby explaining how someone without programming experience could arrive at the casual declaration of \enquote{Now I understand!} upon observing source code.

Defining related constructs is crucial for distinguishing them effectively. 
That is why we introduced a definition for source code comprehension and a conceptual model for empirical investigation based on the surveyed code comprehension theory of the past decades.
Both researchers and the meta-scientific discourse can benefit from our definition and conceptual model, as we have demonstrated by providing and discussing concrete examples.

Our vision for the way forward is as follows.
The individual links in our conceptual model for the design of code comprehension experiments represent their own established research streams that still require more theory.
For example, the community should further investigate how the cognitive comprehension process occurs during the transition between mental states.
Additionally, an evidence-based theory is needed to understand how a complex network of experimental variables influences the formation of mental models.
While both of these research streams have been conducted in the past and are currently gaining momentum again, only now can they be integrated into a cohesive framework at the level of our model.

Finally, construct validity does not end with face and content validity, and many code comprehension studies do not require the development of their own test instruments (especially in correlational studies without treatments). Psychometrically validated tests for code comprehension, built upon the introduced definition, would be useful in such cases.

\section*{Acknowledgments}

Many years of lively exchange with friends and colleagues from the research community led in the end to the clarity that was necessary to write this paper. I would especially like to thank Stefan Wagner, Sven Apel, and Lena Jäger for their valuable input. Wyrich's work is supported by the European Union as part of the ERC Advanced Grant 101052182.

\bibliographystyle{ACM-Reference-Format}
\bibliography{main}

\end{document}